\documentclass[]{fairmeta}

\usepackage{multirow}
\usepackage{color}
\usepackage{colortbl}
\usepackage{mdframed}
\usepackage{caption}
\usepackage{subcaption}
\usepackage{algpseudocode}
\usepackage{algorithm}
\usepackage{amsthm}
\usepackage{amssymb}
\newtheorem*{remark}{Remark} 

\newtheorem{theorem}{Theorem}[section]

\title{External Large Foundation Model: How to Efficiently Serve Trillions of Parameters for Online Ads Recommendation}

\author[1, *]{Mingfu Liang}
\author[1, *]{Xi Liu}
\author[1]{Rong Jin}
\author[1]{Boyang Liu}
\author[1]{Qiuling Suo}
\author[1]{Qinghai Zhou}
\author[1]{Song Zhou}
\author[1]{Laming Chen}
\author[1]{Hua Zheng}
\author[1]{Zhiyuan Li}
% ...

\affiliation[1]{AI at Meta}
\affiliation[2]{University of Illinois Urbana-Champaign}

\contribution[*]{Both authors contributed equally to the paper}

\abstract{Ads recommendation is a prominent service of online advertising systems and has been actively studied. Recent studies indicate that scaling-up and advanced design of the recommendation model can bring significant performance improvement. However, with a larger model scale, such prior studies have a significantly increasing gap from industry as they often neglect two fundamental challenges in industrial-scale applications. 
First, training and inference budgets are restricted for the model to be served, exceeding which may incur latency and impair user experience. 
Second, large-volume data arrive in a streaming mode with data distributions dynamically shifting, as new users/ads join and existing users/ads leave the system. 
We propose the External Large Foundation Model (ExFM) framework to address the overlooked challenges. 
Specifically, we develop external distillation and a data augmentation system (DAS) to control the computational cost of training/inference while maintaining high performance. 
We design the teacher in a way like a foundation model (FM) that can serve multiple students as vertical models (VMs) to amortize its building cost. 
We propose Auxiliary Head and Student Adapter to mitigate the data distribution gap between FM and VMs caused by the streaming data issue.
Comprehensive experiments on internal industrial-scale applications and public datasets demonstrate significant performance gain by ExFM.}

\date{\today}

\begin{document}

\maketitle

\section{Introduction}

Ads recommendation is an important service provided by online advertising systems, whose model performance can impact user experience. It has been actively studied to enhance model performance by advanced designs~\cite{zhu2024collaborative,zhang2022dhen,geng2022recommendation,wang2021dcn, li2020interpretable,ying2018graph} and scaling-up model complexity~\cite{zhang2024wukong,pan2024ads,anil2022factory}, especially after observing the remarkable performance of trillion-parameter models such as GPT-4~\cite{achiam2023gpt}, LLaMa~\cite{touvron2023llama}, etc.
\
However, prior studies often neglect two fundamental challenges in industrial-scale applications:
\begin{itemize}
    \item{C1}: Restricted training and inference latency for the serving models. Industrial platforms usually have to recursively update the models, and evaluate the prediction score of O(100)$\sim$O(100K) ads for each request within a restricted latency. Exceeding latency may degrade model performance and impair user experience. 
    \item{C2}: Large-volume streaming data arrive with data distributions dynamically shifting. This is largely due to the evolving nature of advertising systems: new users/ads join, and existing users/ads leave the system as time passes. This implies that multi-pass training would risk over-fitting and is the most salient difference between industry and academia.
\end{itemize}

\begin{figure*}[htbp]
  \centering
  \includegraphics[width=\textwidth]{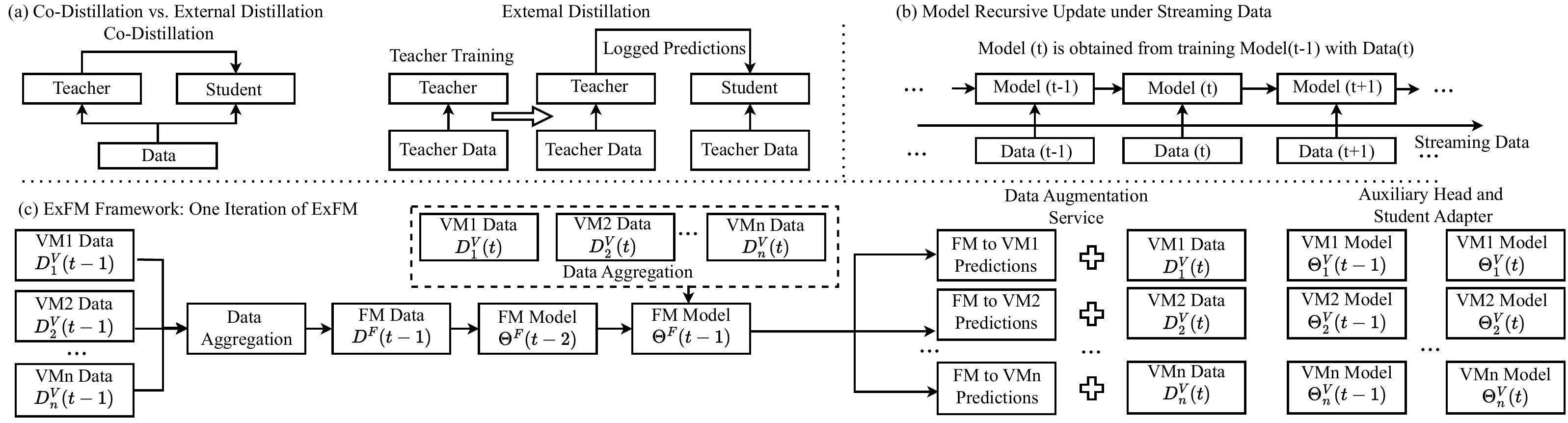}
  \caption{(a) Co-Distillation vs. External Distillation; (b) Model update under streaming data; (c) Our proposed ExFM framework that enabled trillions-parameter model serving with newly designed data augmentation system~(DAS) and external distillation; The proposed auxiliary head and student adapter are developed to mitigate data distribution gap between FM and VMs.}
  \label{fig:ExFM}
\end{figure*}

To overcome the restricted inference latency challenge in C1, multiple prior studies employ knowledge distillation (KD)~\cite{liu2022position, kang2021item,kweon2021bidirectional,zhu2020ensembled}. 
As shown by Figure~\ref{fig:ExFM}(a), a large teacher model is co-trained with a compact student model to be used for serving. 
The teacher's knowledge is continuously distilled into the student. 
Unfortunately, the co-training will increase the training computational cost of the serving model, which cannot meet the industrial restriction on the training latency of the serving model. 
Besides, ads recommendation often involves multiple serving models, each corresponding to one specific service or ranking stage (e.g., early or later/ranking stage). 
Building and maintaining a dedicated large teacher model for each serving model is too inefficient and non-scalable. 
To sum up, the ideal teacher model (1) is separately trained with the student model, i.e., external distillation, and (2) is 1-to-N, able to benefit multiple student models, like a Foundation Model (FM).

\begin{figure*}[htbp]
    \centering
    \includegraphics[width=0.6\linewidth]{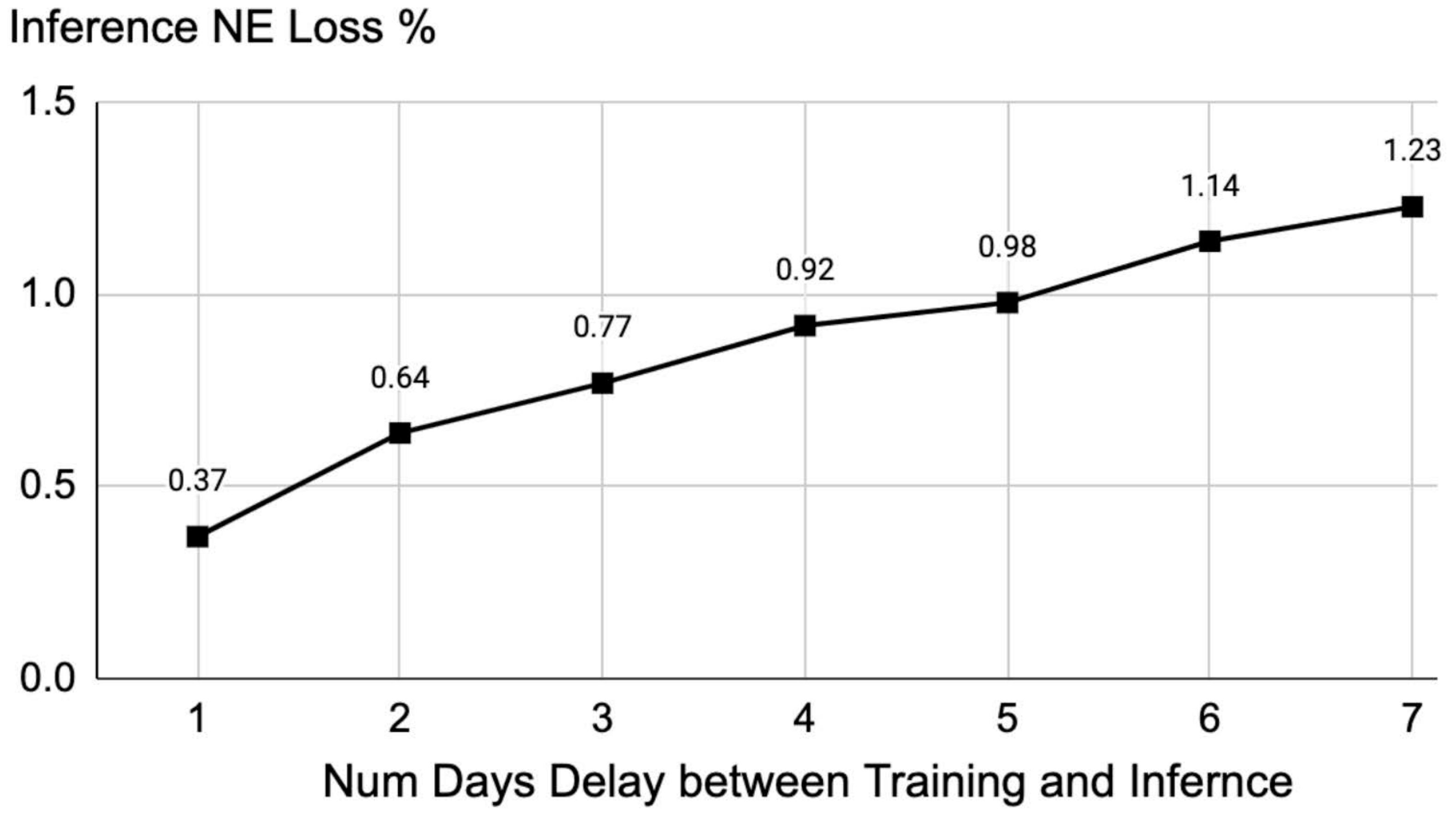}
    \caption{Click Through Rate~(CTR) prediction on the internal dataset. The bigger the staleness, the larger NE Loss.}

    \label{fig:staleness}
\end{figure*}

The streaming and non-stationary nature of recommendation data in C2 makes it challenging to realize the ideal teacher model. 
As shown by Figure~\ref{fig:ExFM}(b), models need to be recursively trained to capture the up-to-date distribution shifting. Figure~\ref{fig:staleness} illustrates an example of performance degradation under model staleness over internal datasets, where the model is trained with 74-billion examples and stops training with new examples since inference. 
Compared to the baseline being trained with new examples daily, its inference \texttt{NE} (normalized entropy)~\cite{he2014practical} loss keeps enlarging as the delay inclines. 
Besides, since the teacher aggregates various students' training data from different services, the distribution of the teacher model's data will be more sophisticated. 
Those factors lead to two potential distribution gaps between the teacher model and the student models: (1) Cross-domain Bias. For an individual student, the teacher may carry bias due to being trained by not only that student's data but an aggregation of all students' data under distribution shifting. (2) Freshness Gap. The teacher predictions used in external distillation are from a teacher with a slight delay w.r.t. training data compared to the student under training.

In light of the above limitations, we propose the \underline{Ex}ternal Large \underline{F}oundation \underline{M}odel (ExFM) framework to address the challenges overlooked by prior studies. 
Figure~\ref{fig:ExFM}(c) presents a big picture of the ExFM framework. 
To avoid additional training or inference computational costs on the serving model, ExFM employs external distillation where the teacher model is separately trained, and the teacher's predictions are offline logged as external supervision for the student training.
To amortize the resources needed for training and maintenance of the teacher model, ExFM aggregates the data from multiple student traffic as the training data of the teacher, like a 1-to-N Foundation Model that can serve multiple students as the vertical models (VMs), i.e. production model. 
To alleviate the Cross-domain Bias transferred from FM to VMs, ExFM proposes Auxiliary Head (AH) that uses an isolated task arch to receive the FM supervision instead of direct on serving head. The simple change is found effective both provably and empirically. 
To mitigate the Freshness Gap between FM and VMs, ExFM proposes Student Adapter (SA) that learns to transform FM predictions before being used as VMs' supervision.

The main contributions of the paper are summarized as follows

\begin{itemize}
\item We re-define the ads recommendation problem to be explored under two fundamental challenges from industrial-scale applications overlooked by prior studies: (1) restricted training and inference budget for serving model, and (2) streaming data under continuously shifting distributions.
\item We propose ExFM framework to overcome above challenges, where (1) the teacher is an FM to cover multiple VMs to amortize the maintaining cost; (2) external distillation~(Sec.~\ref{das}), not co-distillation, is designed with the system to avoid training or inference cost to models to be served (i.e., VM); (3) Auxiliary Head~(Sec.~\ref{ah}), i.e., an isolated task instead of VM's serving head, is used to process FM' supervision, provably alleviating the bias transfer from FM to VMs, (4) Student Adapter~(Sec.~\ref{sa}) with specific learning algorithms is proposed to reduce the staleness gap between FM and VMs, with mathematical guarantee.
\item We conducted extensive experiments~(Sec.~\ref{experiments}) on both internal industrial-scale datasets and public datasets and demonstrated that ExFM can enable efficient serving of a trillion-parameter model. We observed promising performance enhancement by employing ExFM on VMs from different domains/tasks/stages. We also provide ablation studies to reveal the impact of hyper-parameters.
\end{itemize}

\section{Problem Formulation}

We re-define the ads recommendation problem to match industrial-scale applications with two overlooked fundamental challenges:

\begin{itemize}
    \item Restricted training and inference latency for serving models. Ideally, the model enhancement should not incur any training or inference computational costs on serving models.
    \item Training and inference data are large volumes of streaming data with continuously shifting data distribution. Models have to keep training on fresh examples to perform well.
\end{itemize}

\textit{The new ads recommendation problem is to predict user interest in ads based on their interaction data with dynamic data distribution shift constrained by restricted training and inference latency}.
Formally, let $\Theta^F(t)$ be the FM parameters after training $\Theta^F(t-1)$ on FM's new data $D^F(t)$ at time $t$, and $\Theta_i^V(t)$ be VM $i$'s parameters after training $\Theta_i^V(t-1)$ on VM's new data $D_i^V(t)$, $i=1,\ldots,n$.

\section{Methods}

\subsection{Overview}
Figure~\ref{fig:ExFM}(c) illustrates the details of the ExFM framework with one model iteration as an example. Specifically, FM and VMs are recursively trained on streaming data $D^F(t)$ and $\{D^V_i(t)\}_{i=1}^{n}$ incrementally arrived at time $t$. To amortize the building and maintenance costs, the teacher model is built as an FM that can make prediction on multiple VMs' traffic. Correspondingly, the training data of FM $D^F(t-1)$ is obtained by aggregating the cross-traffic VM's training data $\{D^V_i(t-1)\}_{i=1}^{n}$. After training FM model $\Theta^F(t-2)$ on $D^F(t-1)$, FM model parameters are updated to $\Theta^F(t-1)$.

To not incur additional training or inference costs to VMs, external distillation, as shown in Figure~\ref{fig:ExFM}(a), is employed. Specifically, Teacher/FM is trained on $D^F(t-1)$ separately from Students/VMs. After training, Teacher/FM inferences on Students'/VMs' training data $\{D^V_i(t)\}_{i=1}^{n}$ to generate predictions as external supervision to Students/VMs. The external supervision, along with the VMs' training data $\{D^V_i(t)\}_{i=1}^{n}$ will be used to train the Students/VMs' models $\{\Theta_i^V(t-1)\}_{i=1}^n$, updating them to $\{\Theta_i^V(t)\}_{i=1}^n$. The model snapshots of Students/VMs $\{\Theta_i^V(t)\}$ are used to serve traffic for the next period $[t, t+1)$. FM's supervision logging and the merging with Students/VMs training data are achieved by Data Augmentation Service (DAS), elaborated in Section~\ref{das}.

Since FM's training data $D^F(t-1)$ is an aggregation of VMs' data $\{D^V_i(t)\}_{i=1}^{n}$ from multiple traffics, for one specific traffic, FM supervision readily carries cross-domain bias. Multiple prior studies handle this problem by carefully designing the distillation losses~\cite{gou2021knowledge}. Motivated by that, we propose a flexible solution that can cover different proposals from prior studies, called Auxiliary Head, which can provably reduce the bias transfer from FM supervision to VM and significantly enhance FM's benefits on VM in experiments. Details are provided in Section~\ref{ah}.

Also note that there is always a delay between FM training and FM inference used in VMs' training: FM $\Theta^F(t-1)$ inferences on data $\{D^V_i(t)\}_{i=1}^{n}$ to update VMs to $\{\Theta_i^V(t)\}_{i=1}^n$. Due to the streaming and distribution shifting nature of the industry data, this delay will create a Freshness Gap between the FM's supervision and the VMs' training and impair the beneficial impact of FM. Student Adapter is proposed to alleviate the gap and elaborated in Section~\ref{sa}.
\begin{figure}[htbp]
    \centering
    \includegraphics[width=\textwidth]{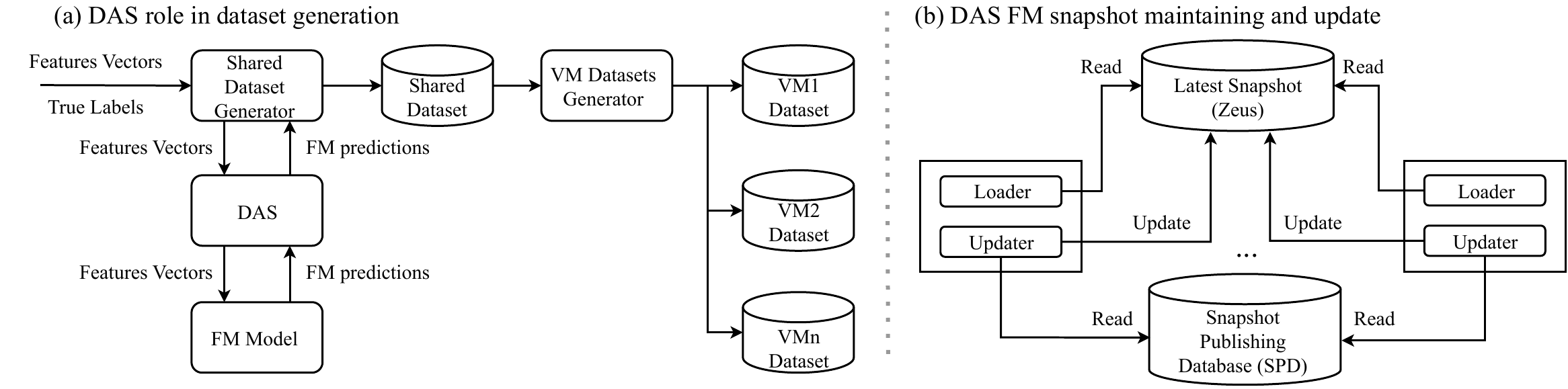} 
    \caption{Data Augmentation Service (DAS)}
    \label{fig:das}
\end{figure}
\subsection{Data Augmentation Service~(DAS)}\label{das}

Data Augmentation Service (DAS) logs FM's supervision on the fly when preparing training data for VMs, fulfilled in two stages. 
As shown by Figure~\ref{fig:das}(a), first, all VMs' features and true labels are joined together into a shared dataset. 
There is a large time window to wait for the feedback (i.e., true label) for this training example (e.g., 5 to 90 minutes for CTR feedback, or 1 day for CVR feedback). 
During this time window, DAS will call FM to get FM's inference on VM's features, which usually has a large latency budget (in seconds level) sufficient to evaluate a large-scale FM. 
Then each VM picks its own datasets from the shared dataset for training. 
This ensures that the inference of FM only needs to cover a tiny percentage of ranked ads in training data, and VMs with overlapped traffic can further amortize the inference cost of the FM. 

As mentioned earlier, models need to be recursively trained to capture the distribution shifting of streaming data. 
Consequently, FM publishes model snapshots regularly. DAS has to actively detect if there is a new snapshot and switch to it if so. 
In practice, it may take several minutes to switch to a snapshot, as loading also takes time. 
Before completeness, we want the old snapshot to still maintain high availability for FM supervision generation. 
To avoid wasting resources, we also want the new snapshot to be loaded only once. 
To this end, DAS separates the FM model snapshot publishing and the latest model snapshot database. 
As shown by Figure~\ref{fig:das}(b), a snapshot is regularly published in the Snapshot Publishing Database (SPD), while the latest snapshot to be loaded for generating supervision is maintained in Zeus, a distributed metadata store based on Apache Zookeeper~\cite{zookeeper} with strong consistency. 
Zeus maintains the snapshot with a lease to do automatic garbage collection of old snapshots.
The existing lease will be extended automatically at a fixed frequency until a new snapshot is discovered. 
For DAS tasks, Zeus is readable and writable, while SPD is only readable. 
The updater of the DAS task regularly queries SPD to detect if there is a new snapshot. 
If a new snapshot is discovered, the updater will read it from SPD and write it into Zeus. 
The loader of the DAS task is responsible for loading the latest snapshot to generate supervision. 
DAS treats all tasks equally instead of having a special primary task to be scalable to distributed settings.

\subsection{Auxiliary Head~(AH)}\label{ah}
\begin{figure}[htbp]
  \centering
  \includegraphics[width=\textwidth]{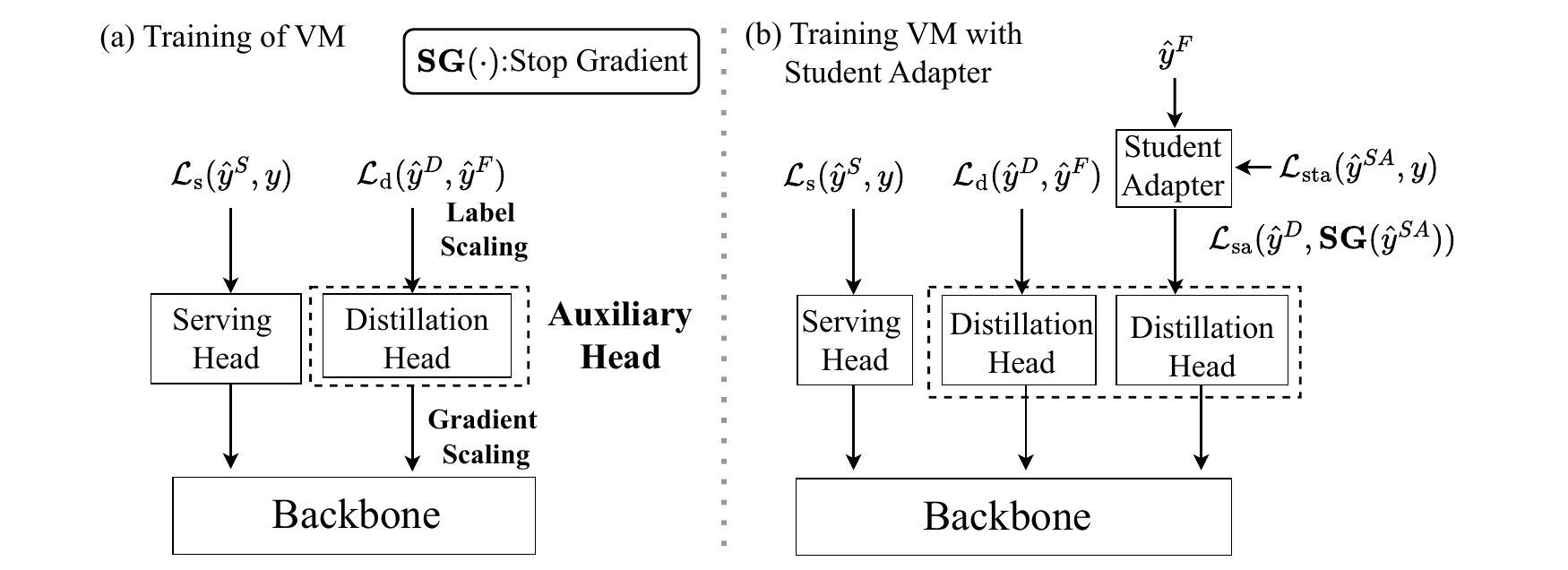}
  \caption{Auxiliary Head~(AH) and Student Adapter~(SA). $\text{SG}$ means to stop the gradient from the output of SA.}
  \label{fig: ah and sa}
\end{figure}

The baseline of knowledge distillation is through label smoothing of the ground-truth label $y$ and the pseudo-label $\hat{y}^{F}$ from FM. 
The model architecture of VM consists of two parts: the backbone and a serving head. 
We define the output vector of the backbone as $\mathbf{x}$, and $\mathbf{x}$ will be sent to the serving head. The output vector of the serving head is defined as $\hat{y}^{S} =\phi(\mathbf{x})$, where $\phi$ denotes the multi-layer perceptron (MLP).
The training objective is binary cross-entropy and can be generally defined as 
\begin{equation}
    \label{bce loss}
    h(\hat{y}, y) = y\cdot \log \left(\sigma(\hat{y})\right)+\left(1-y\right) \cdot \log \left(1-\sigma(\hat{y})\right),
\end{equation}
where $\sigma(x)=1/(1+\text{exp}(-x))$ denotes the Sigmoid activation function and $\hat{y}$ denotes the prediction. 
Accordingly, the training loss of VM with knowledge distillation is defined as
\begin{equation}
    \label{distillation loss}
    \mathcal{L}_{\text{kd}}(\hat{y}^{S},\hat{y}^{F},y) = h(\hat{y}^{S}, y) + h(\hat{y}^{S}, \hat{y}^{F})
\end{equation}
where $\hat{y}^{S}$ denotes the output of the serving head.
The gradients with respect to $y$ and $\hat{y}^{F}$ are entangled in the serving head and flowing back to the backbone.

Using a single head to consume both labels may induce bias from FM to VM. 
Therefore, we propose the Auxiliary Head ($\text{AH}$) that leverages a separate head to consume the pseudo-label from FM.
In this case, the model architecture of VM consists of the backbone, the serving head, and a distillation head as $\text{AH}$.
The output of the backbone $\mathbf{x}$ will be sent to different heads, and the output of $\text{AH}$ is from the distillation head $\hat{y}^{D}$, where $\hat{y}^{D}$ is defined as $\hat{y}^{D} = \psi(\mathbf{x})$. $\psi$ represents an additional MLP head which is built on top of the backbone.
By disentangling the serving head and distillation loss, the serving head is only supervised by the ground-truth label $y$ as 
\begin{equation}
    \label{serving head loss}
    \mathcal{L}_{\text{s}}(\hat{y}^{S}, y) = h(\hat{y}^{S}, y).
\end{equation}
The distillation head is supervised by the pseudo-label $\hat{y}^{F}$ as 
\begin{equation}
    \label{distllation head loss}
    \mathcal{L}_{\text{d}}(\hat{y}^{D}, \hat{y}^{F}) = h(\hat{y}^{D}, \hat{y}^{F}).
\end{equation}

We further provide the theoretical insight that, compared to adding the auxiliary loss to the serving task for distillation, we prove that integrating the auxiliary head for knowledge distillation~(KD) can alleviate the bias transfer from FM to VMs. 

\vspace{5pt}
\begin{mdframed}[backgroundcolor=gray!8]
\begin{minipage}{\linewidth}
\begin{theorem}\label{theo: KD by AH}
(Informal)~The VM will contain bias from the FM when the VM is trained with KD by a single serving head. In contrast, the VM is guaranteed to find the optimal solution for predicting $y^{gt}$ with KD by auxiliary heads.
\end{theorem}
\end{minipage}
\end{mdframed}

\begin{remark}
Calibration is one of the most critical metric in ads recommendation system. When we perform knowledge distillation with single serving head, the introduced bias will lead to mis-calibration issue, while separate head can guarantee the calibration will not be effected. More specifically, if the teacher model is mis-calibrated, the bias transfer from the teacher model from a single serving head can lead to mis-calibration in the student model with high probability. In contrast, distillation with an auxiliary head mitigates this issue by reducing the propagation of teacher model's bias through the separated serving head.
\end{remark}

The proof of Theorem~\ref{theo: KD by AH} in the Appendix Sec.~\ref{proof of theorem for AH}, where we prove by construction using a two-layer linear model to show that (1) the bias in the pseudo labels from the FM is partially compensated by the separated prediction head for KD, and (2) fewer ``biased'' gradients will flow from the prediction head for KD to the shared backbone, leading to a better-generalized performance than the single serving head for KD.

However, only isolating the supervision from $\hat{y}^{F}$ to $y$ is not enough to bring effective knowledge transfer in practice. 
This is due to the issues originating from the intrinsic properties of the training data of Ads engagement, i.e., the majority of engagement data does not convert to actions like clicking; thus, the FM primarily predicts majority of $\hat{y}^{F}$ close to zero, making the KD task fitting to a difficult long-tailed distribution. These practical issues hinder the effectiveness of the KD with AH. 

To remedy the above issue, we propose amplifying the FM's distillation effect by \textbf{Gradient Scaling~(GS), Label Scaling~(LS), and Loss Weighting~(LW)}. 
The gradient scaling scales the gradient of the distillation head to backbone with a hyperparameter $\beta$ during training to enlarge the impact of the distillation from the FM.

For label scaling, we multiply the $\hat{y}^{F}$ with a constant $\alpha$ to enlarge its magnitude and clip the $\alpha \cdot \hat{y}^{F}$ correspondingly to avoid the scaled label being out-of-bound, e.g., $\alpha \cdot \hat{y}^{FM}$ needs to be smaller or equal to 1. 
Lastly, we apply the loss weighting to $\mathcal{L}_{\text{d}}$ with a hyperparameter $w$. Accordingly, the final training loss $\mathcal{L}_{\text{ah}}$ for VM  with AH is:
\begin{equation}
    \label{auxiliary head loss}
    \mathcal{L}_{\text{ah}}=\mathcal{L}_{\text{s}}(\hat{y}^{S}, y) + w * \mathcal{L}_{\text{d}}(\hat{y}^{D}, \alpha \cdot \hat{y}^{F})
\end{equation}

\subsection{Student Adapter~(SA)}\label{sa}
To further alleviate the bias from the $\hat{y}^{F}$, we propose to provide additional distillation using the adapted $\hat{y}^{F}$ based on $y$, called Student Adapter~($\text{SA}$). The output of $\text{SA}$ is defined as $\hat{y}^{SA} = \text{MLP}(\hat{y}^{F})$. As shown in Figure~\ref{fig: ah and sa}~(b), SA is trained by the loss
\begin{equation}
    \label{student adapter training loss}
    \mathcal{L}_{\text{sta}}(\hat{y}^{SA}, y) = h(\hat{y}^{SA}, y)
\end{equation}
When training VM with $\text{SA}$, we first optimize $\mathcal{L}_{\text{sta}}(\hat{y}^{SA}, y) $ to update $\text{SA}$ and get $\hat{y}^{SA}$. To use $\hat{y}^{SA}$ for training VM, we stop the gradient of $\hat{y}^{SA}$, i.e., $\text{SG}(\hat{y}^{SA})$, to avoid the gradient of VM flowing back to $\text{SA}$, and we consume $\text{SG}(\hat{y}^{SA})$ with another distillation head to update VM with the following loss 
\begin{equation}
    \label{student adapter loss for VM}
    \mathcal{L}_{\text{sa}}(\hat{y}^{D}, \text{SG}(\hat{y}^{SA})) = h(\hat{y}^{D}, \text{SG}(\hat{y}^{SA})).
\end{equation}
The algorithm is detailed in Algorithm~\ref{alg: 1}.

We provide theoretical insight that SA can mitigate the freshness gap between FM and VM by considering the standard regression problem for analysis. The proof is provided in Appendix~\ref{proof of theorem 3.2}.

\vspace{5pt}
\begin{mdframed}[backgroundcolor=gray!8]
\begin{minipage}{\linewidth}
\begin{theorem}\label{theo: KD by student adapter}
With a probability $1-\delta$, the underlying regression model $w$ and the optimal solution $w_{V}$ achieved by training VM with Student Adapter (SA) is bounded as:
$$ \left|w_{V}-w\right|_2 \leq O\left(\gamma \sqrt{\frac{(d s)^{1 / 2}}{N} \log \frac{1}{\delta}}\right),$$
where $N$ denotes the number of training samples, $d$ denotes the input dimension, $\gamma$ is a constant, $s$ denotes the model dimension.
 \end{theorem}
\end{minipage}
\end{mdframed}

Compared to the bound for the standard regression model when training with KD, the bound is reduced by the factor $(d/s)^{1/4}$, implying that if the distribution is changing on the time scale between $d$ and $\sqrt{ds}$, the student adapter can quickly catch up the drift.

\begin{algorithm}[th]
\caption{Train VM with Student Adapter for one iteration}
\raggedright
\textbf{Input:} Pseudo-label $\hat{y}^{F}$ from FM, ground-truth label $y$, learnable model parameter $\Theta^{SA}$ of the Student Adapter $\text{SA}$, learnable parameters $\Theta^{V}$ of the VM.

\textbf{Output:} Model parameter of the Student Adapter $\Theta^{SA}$, and the model parameter of VM $\Theta^{V}$

\begin{algorithmic}[1]
\State Optimize $\mathcal{L}_{\text{sta}}(\hat{y}^{SA}, y)$~(Eqn.~\ref{student adapter training loss}) to update $\Theta^{SA}$
\State Get $\text{SG}(\hat{y}^{SA})$ by stopping gradient of the output for $\text{SA}$
\State Optimize $\mathcal{L}_{\text{ah}}~(Eqn.~\ref{auxiliary head loss}) + \mathcal{L}_{\text{sa}}$~(Eqn.~\ref{student adapter loss for VM}) to update $\Theta^{V}$. 
\State \textbf{Return $\Theta^{SA}$ and $\Theta^{V}$}
\end{algorithmic}
\label{alg: 1}
\end{algorithm}

\begin{table}[htbp]
  \centering
  \caption{[Public] Dataset Statistics}
    \begin{tabular}{ccc}
    \toprule
    Dataset & \#Samples & \#Features \\
    \midrule
    Amazon Electronics & 3.0M  & 6 \\
    TaobaoAd & 25.0M & 22 \\
    KuaiVideo & 13.7M & 9 \\
    \bottomrule
    \end{tabular}%
  \label{tab: dataset stat}%
\end{table}%

\section{Experiments}
\label{experiments}
We evaluate the proposed ExFM on both internal industrial-scale and public datasets to answer the following research questions:
\begin{itemize}
    \item{Q1}: How effective is the proposed ExFM in elevating the performance of a single VM and multiple VMs, e.g., from different tasks, domains or stages?~(Sec.~\ref{Effectiveness Results})
    \item{Q2}: How effective are the proposed Auxiliary Head and Student Adapter?~(Sec.~\ref{sec:delta of ah/sa})
    \item{Q3}: How do the values of hyperparameters impact the ExFM performance?~(Sec.~\ref{sec: parameter analysis})
\end{itemize}
For better readability, each part of discussion will first cover internal datasets, then public datasets. The caption of each figure and table starts from [Internal] if obtained on internal datasets and [Public] on public datasets.

\subsection{Experiment Setup}

\subsubsection{\textbf{Datasets}}
\label{Datasets}
We conduct comprehensive experiments on internal industrial-scale datasets with billions of training examples, from ads CTR prediction tasks. Each training example contains O(1k) features for a user and ad pair, and various types of feedback such as click, post-click conversion, etc. Training examples of individual VMs come from the corresponding services or stages and may contain service-dedicated features and feedback. Training examples of the FM aggregate the training examples of VMs.
Billions of training examples arrive daily for both VMs and FM.
All training is one-pass, and training batches are ordered by arrival time.

For public datasets, we use TaobaoAd~\cite{Tianchi}, Amazon Electronics~\cite{he2016ups} and KuaiVideo~\cite{Kuaishou} as they are widely adopted and have the timestamp feature that facilitates our experiments for streaming setting. The dataset statistics are briefly summarized in Table~\ref{tab: dataset stat}. For the streaming setting, we split each dataset into 8 splits (days) and divide them into the FM train set, VM train set, and VM test set based on the UNIX timestamps. For example, FM is trained on the first 4 days' data and inference on 5th day's data to generate supervision. VM is trained on the 5th day with FM's supervision and inference on the 6th day to get the AUC/LogsLoss reading. We can repeat this process for 3 times until Day 8.
On TaobaoAd and Amazon Electronics, we train FM and VMs on the CTR prediction task. 
To validate FM's impact on different-domain VMs, we split TaobaoAd dataset by the value of the categorical feature `new\_user\_class\_level'. The original `new\_user\_class\_level' feature has five levels; we combine them into three levels as domains to ensure that each domain has almost the same training data size. 
To experiment the FM's impact on different-task VMs, we use KuaiVideo~\cite{Kuaishou} due to its various feedbacks. We use prediction of `Click,' `Follow,' and `Like,' as three different tasks, each of which has a dedicated VM to handle. 

\begin{table*}[htbp]
  \centering
  \footnotesize
  \caption{[Public] ExFM performance under different FM/VM, AH/SA and datasets choices. We choose DMIN~\cite{xiao2020deep}, DCN~\cite{wang2017deep}, and DCNv2~\cite{wang2021dcn} as FM, and FaM~\cite{rendle2010factorization}, FmFaM~\cite{sun2021fm2} and DeepFaM~\cite{guo2017deepfm} as VM.}
    \begin{tabular}{cccccccc}
    \toprule 
    & \multicolumn{1}{c}{Dataset $\longrightarrow$} & \multicolumn{6}{c}{TaobaoAd}                     \\
\cmidrule{3-8}          & \multicolumn{1}{c}{Vertical Models $\longrightarrow$} & \multicolumn{2}{c}{FaM} & \multicolumn{2}{c}{FmFaM} & \multicolumn{2}{c}{DeepFaM} \\
\cmidrule{3-8}    Teacher Model & \multicolumn{1}{c}{Methods} & AUC  & LogLoss & AUC   & LogLoss & AUC   & LogLoss \\
\midrule
    \multicolumn{1}{c}{\multirow{4}[0]{*}{DMIN}} & w/o distill & 0.5389 & 0.7228 & 0.5477 & 0.5388 & 0.5488 & 0.5713  \\
          & distill w/ $\mathcal{L}_{\text{kd}}$ & 0.5408 & 0.6484 & 0.5482 & 0.5349 & 0.5576 & 0.2943  \\
          & \cellcolor[rgb]{ .816,  .816,  .816}distill w/ $\mathcal{L}_{\text{ah}}$ & \cellcolor[rgb]{ .816,  .816,  .816}0.5449 & \cellcolor[rgb]{ .816,  .816,  .816}0.6451 & \cellcolor[rgb]{ .816,  .816,  .816}0.5494 & \cellcolor[rgb]{ .816,  .816,  .816}0.5223 & \cellcolor[rgb]{ .816,  .816,  .816}0.5630 & \cellcolor[rgb]{ .816,  .816,  .816}0.2885 \\
          & \cellcolor[rgb]{ .816,  .816,  .816}distill w/ $\mathcal{L}_{\text{ah}}+\mathcal{L}_{\text{sa}}$  & \cellcolor[rgb]{ .816,  .816,  .816}\textbf{0.5462} & \cellcolor[rgb]{ .816,  .816,  .816}\textbf{0.6411} & \cellcolor[rgb]{ .816,  .816,  .816}\textbf{0.5521} & \cellcolor[rgb]{ .816,  .816,  .816}\textbf{0.5208} & \cellcolor[rgb]{ .816,  .816,  .816}\textbf{0.5648} & \cellcolor[rgb]{ .816,  .816,  .816}\textbf{0.2871} \\
\midrule
    \multicolumn{1}{c}{\multirow{4}[0]{*}{DCN}} & w/o distill & 0.5389 & 0.7228 & 0.5477 & 0.5388 & 0.5488 & 0.5713  \\
          & distill w/ $\mathcal{L}_{\text{kd}}$ & 0.5397 & 0.6587 & 0.5479 & 0.5359 & 0.5556 & 0.3183\\
          & \cellcolor[rgb]{ .816,  .816,  .816}distill w/ $\mathcal{L}_{\text{ah}}$ & \cellcolor[rgb]{ .816,  .816,  .816}0.5438 & \cellcolor[rgb]{ .816,  .816,  .816}0.6565 & \cellcolor[rgb]{ .816,  .816,  .816}0.5495 & \cellcolor[rgb]{ .816,  .816,  .816}0.5228 & \cellcolor[rgb]{ .816,  .816,  .816}0.5612 & \cellcolor[rgb]{ .816,  .816,  .816}0.2918 \\
          & \cellcolor[rgb]{ .816,  .816,  .816}distill w/ $\mathcal{L}_{\text{ah}}+\mathcal{L}_{\text{sa}}$ & \cellcolor[rgb]{ .816,  .816,  .816}\textbf{0.5458} & \cellcolor[rgb]{ .816,  .816,  .816}\textbf{0.6449} & \cellcolor[rgb]{ .816,  .816,  .816}\textbf{0.5517} & \cellcolor[rgb]{ .816,  .816,  .816}\textbf{0.5213} & \cellcolor[rgb]{ .816,  .816,  .816}\textbf{0.5624} & \cellcolor[rgb]{ .816,  .816,  .816}\textbf{0.2901} \\
\midrule
    \multicolumn{1}{c}{\multirow{4}[0]{*}{DCNv2}} & w/o distill & 0.5389 & 0.7228 & 0.5477 & 0.5388 & 0.5488 & 0.5713\\
          & distill w/ $\mathcal{L}_{\text{kd}}$ & 0.5401 & 0.6580 & 0.5481 & 0.5351 & 0.5561 & 0.3119\\
          & \cellcolor[rgb]{ .816,  .816,  .816}distill w/ $\mathcal{L}_{\text{ah}}$ & \cellcolor[rgb]{ .816,  .816,  .816}0.5440 & \cellcolor[rgb]{ .816,  .816,  .816}0.6538 & \cellcolor[rgb]{ .816,  .816,  .816}0.5498 & \cellcolor[rgb]{ .816,  .816,  .816}0.5219 & \cellcolor[rgb]{ .816,  .816,  .816}0.5626 & \cellcolor[rgb]{ .816,  .816,  .816}0.2899 \\
          & \cellcolor[rgb]{ .816,  .816,  .816}distill w/ $\mathcal{L}_{\text{ah}}+\mathcal{L}_{\text{sa}}$ & \cellcolor[rgb]{ .816,  .816,  .816}\textbf{0.5463} & \cellcolor[rgb]{ .816,  .816,  .816}\textbf{0.6401} & \cellcolor[rgb]{ .816,  .816,  .816}\textbf{0.5525} & \cellcolor[rgb]{ .816,  .816,  .816}\textbf{0.5199} & \cellcolor[rgb]{ .816,  .816,  .816}\textbf{0.5639} & \cellcolor[rgb]{ .816,  .816,  .816}\textbf{0.2879}\\
    \bottomrule
    \end{tabular}%
  \label{tab-1: different FM and VM}%
\end{table*}%

\subsubsection{\textbf{Metrics}}
\label{Metrics and Objectives}
We adopt three widely-used evaluation metrics:
\begin{itemize}
     \item \texttt{NE} (normalized entropy)~\cite{he2014practical}, is the \texttt{LogLoss} in Eqn.~\ref{bce loss} normalized by the entropy of the average empirical CTR of the training dataset, providing a data-insensitive evaluation.
     Lower the better, and an NE gain of 0.02\% is considered significant for experiments on internal datasets.
    \item \texttt{AUC} is an aggregate measure of model performance in correctly classifying positive and negative samples across all thresholds, used for public datasets. Higher the better.
    \item \texttt{LogLoss} (cross-entropy loss) measures the distance between model prediction $\hat{y}$ and actual label $y$ by Eqn.~\ref{bce loss}, which will be used for public datasets. Lower the better.
\end{itemize}

\subsubsection{\textbf{Model Arch}}
\label{exp: baselines}
ExFM is model-agnostic and our purpose is to show the delta impact of FM on VM. In our problem setting, FM is assumed to have larger capacity and demand more computation resource for training and inference. To be consistent, FM we experimented with on internal datasets is often 15 to 600 times larger than the VM. FM enjoys more advanced and resource intensive model architecture such as an internal version of Interformer~\cite{zeng2024interformer}, SUM\cite{zhang2024scaling}, Wukong~\cite{zhang2024wukong}, DHEN~\cite{zhang2022dhen}, etc., while VM adopts a shrunk version of FM or less resource intensive structure such as DLRM~\cite{naumov2019deep}.

On public datasets, we use the models that attain SOTA performance on TaobaoAd, Amazon Electronics, and KuaiVideos as FMs based on BARS~\cite{DBLP:conf/sigir/ZhuDSMLCXZ22} benchmark. Specifically, we use DMIN~\cite{xiao2020deep}, DCN~\cite{wang2017deep}, and DCNv2~\cite{wang2021dcn} as different variants of FM. Regarding VMs, Factorization Machine~(FaM)~\cite{rendle2010factorization} and its two variants, i.e., FmFaM~\cite{sun2021fm2} and DeepFaM~\cite{guo2017deepfm} are considered, as they are less complex and low latency to SOTA.
For instance, the parameter of FaM is 4.23M on Amazon Electronics while DMIN is 5.94M, with a near 30\% parameter decrease. Our purpose is to study the delta impact of FM on VMs under different choices of the two.
More details of the models and training settings are in Appendix~\ref{Details of datasets, models, and training configurations}.

\begin{table*}[htbp]
  \centering
  \footnotesize
  \caption{[Public] ExFM performance under different FM/VM, AH/SA and datasets choices. We choose DMIN~\cite{xiao2020deep}, DCN~\cite{wang2017deep}, and DCNv2~\cite{wang2021dcn} as FM, and FaM~\cite{rendle2010factorization}, FmFaM~\cite{sun2021fm2} and DeepFaM~\cite{guo2017deepfm} as VM.}
    \begin{tabular}{cccccccc}
    \toprule 
    & \multicolumn{1}{c}{Dataset $\longrightarrow$} & \multicolumn{6}{c}{Amazon Electronics} \\
\cmidrule{3-8}          & \multicolumn{1}{c}{Vertical Models $\longrightarrow$} & \multicolumn{2}{c}{FaM} & \multicolumn{2}{c}{FmFaM} & \multicolumn{2}{c}{DeepFaM} \\
\cmidrule{3-8}    Teacher Model & \multicolumn{1}{c}{Methods} & AUC  & LogLoss & AUC   & LogLoss & AUC   & LogLoss \\
\midrule
    \multicolumn{1}{c}{\multirow{4}[0]{*}{DMIN~\cite{xiao2020deep}}} & w/o distill & 0.8331 & 0.5493 & 0.8491 & 0.5311 & 0.8502 & 0.5303 \\
          & distill w/ $\mathcal{L}_{\text{kd}}$ & 0.8349 & 0.5473 & 0.8501 & 0.5305 & 0.8519 & 0.5283 \\
          & \cellcolor[rgb]{ .816,  .816,  .816}distill w/ $\mathcal{L}_{\text{ah}}$ &  \cellcolor[rgb]{ .816,  .816,  .816}0.8399 & \cellcolor[rgb]{ .816,  .816,  .816}0.5431 & \cellcolor[rgb]{ .816,  .816,  .816}0.8538 & \cellcolor[rgb]{ .816,  .816,  .816}0.5232 & \cellcolor[rgb]{ .816,  .816,  .816}0.8551 & \cellcolor[rgb]{ .816,  .816,  .816}0.5165 \\
          & \cellcolor[rgb]{ .816,  .816,  .816}distill w/ $\mathcal{L}_{\text{ah}}+\mathcal{L}_{\text{sa}}$  &  \cellcolor[rgb]{ .816,  .816,  .816}\textbf{0.8418} & \cellcolor[rgb]{ .816,  .816,  .816}\textbf{0.5379} & \cellcolor[rgb]{ .816,  .816,  .816}\textbf{0.8557} & \cellcolor[rgb]{ .816,  .816,  .816}\textbf{0.5141} & \cellcolor[rgb]{ .816,  .816,  .816}\textbf{0.8573} & \cellcolor[rgb]{ .816,  .816,  .816}\textbf{0.5119} \\
\midrule
    \multicolumn{1}{c}{\multirow{4}[0]{*}{DCN~\cite{wang2017deep}}} & w/o distill & 0.8331 & 0.5493 & 0.8491 & 0.5311 & 0.8502 & 0.5303 \\
          & distill w/ $\mathcal{L}_{\text{kd}}$ & 0.8339 & 0.5486 & 0.8497 & 0.5288 & 0.8514 & 0.5290 \\
          & \cellcolor[rgb]{ .816,  .816,  .816}distill w/ $\mathcal{L}_{\text{ah}}$ &  \cellcolor[rgb]{ .816,  .816,  .816}0.8374 & \cellcolor[rgb]{ .816,  .816,  .816}0.5451 & \cellcolor[rgb]{ .816,  .816,  .816}0.8534 & \cellcolor[rgb]{ .816,  .816,  .816}0.5239 & \cellcolor[rgb]{ .816,  .816,  .816}0.8548 & \cellcolor[rgb]{ .816,  .816,  .816}0.5181 \\
          & \cellcolor[rgb]{ .816,  .816,  .816}distill w/ $\mathcal{L}_{\text{ah}}+\mathcal{L}_{\text{sa}}$ & \cellcolor[rgb]{ .816,  .816,  .816}\textbf{0.8401} & \cellcolor[rgb]{ .816,  .816,  .816}\textbf{0.5390} & \cellcolor[rgb]{ .816,  .816,  .816}\textbf{0.8551} & \cellcolor[rgb]{ .816,  .816,  .816}\textbf{0.5153} & \cellcolor[rgb]{ .816,  .816,  .816}\textbf{0.8570} & \cellcolor[rgb]{ .816,  .816,  .816}\textbf{0.5115} \\
\midrule
    \multicolumn{1}{c}{\multirow{4}[0]{*}{DCNv2~\cite{wang2021dcn}}} & w/o distill & 0.8331 & 0.5493 & 0.8491 & 0.5311 & 0.8502 & 0.5303 \\
          & distill w/ $\mathcal{L}_{\text{kd}}$ & 0.8343 & 0.5480 & 0.8503 & 0.5298 & 0.8522 & 0.5272 \\
          & \cellcolor[rgb]{ .816,  .816,  .816}distill w/ $\mathcal{L}_{\text{ah}}$ & \cellcolor[rgb]{ .816,  .816,  .816}0.8392 & \cellcolor[rgb]{ .816,  .816,  .816}0.5441 & \cellcolor[rgb]{ .816,  .816,  .816}0.8541 & \cellcolor[rgb]{ .816,  .816,  .816}0.5225 & \cellcolor[rgb]{ .816,  .816,  .816}0.8568 & \cellcolor[rgb]{ .816,  .816,  .816}0.5131 \\
          & \cellcolor[rgb]{ .816,  .816,  .816}distill w/ $\mathcal{L}_{\text{ah}}+\mathcal{L}_{\text{sa}}$  & \cellcolor[rgb]{ .816,  .816,  .816}\textbf{0.8411} & \cellcolor[rgb]{ .816,  .816,  .816}\textbf{0.5410} & \cellcolor[rgb]{ .816,  .816,  .816}\textbf{0.8563} & \cellcolor[rgb]{ .816,  .816,  .816}\textbf{0.5139} & \cellcolor[rgb]{ .816,  .816,  .816}\textbf{0.8582} & \cellcolor[rgb]{ .816,  .816,  .816}\textbf{0.5104} \\
    \bottomrule
    \end{tabular}%
  \label{tab-2: different FM and VM}%
\end{table*}%

\subsection{Effectiveness of FM on VMs} \label{Effectiveness Results}

\begin{figure}[htbp]
    \centering
    \includegraphics[width=0.6\linewidth]{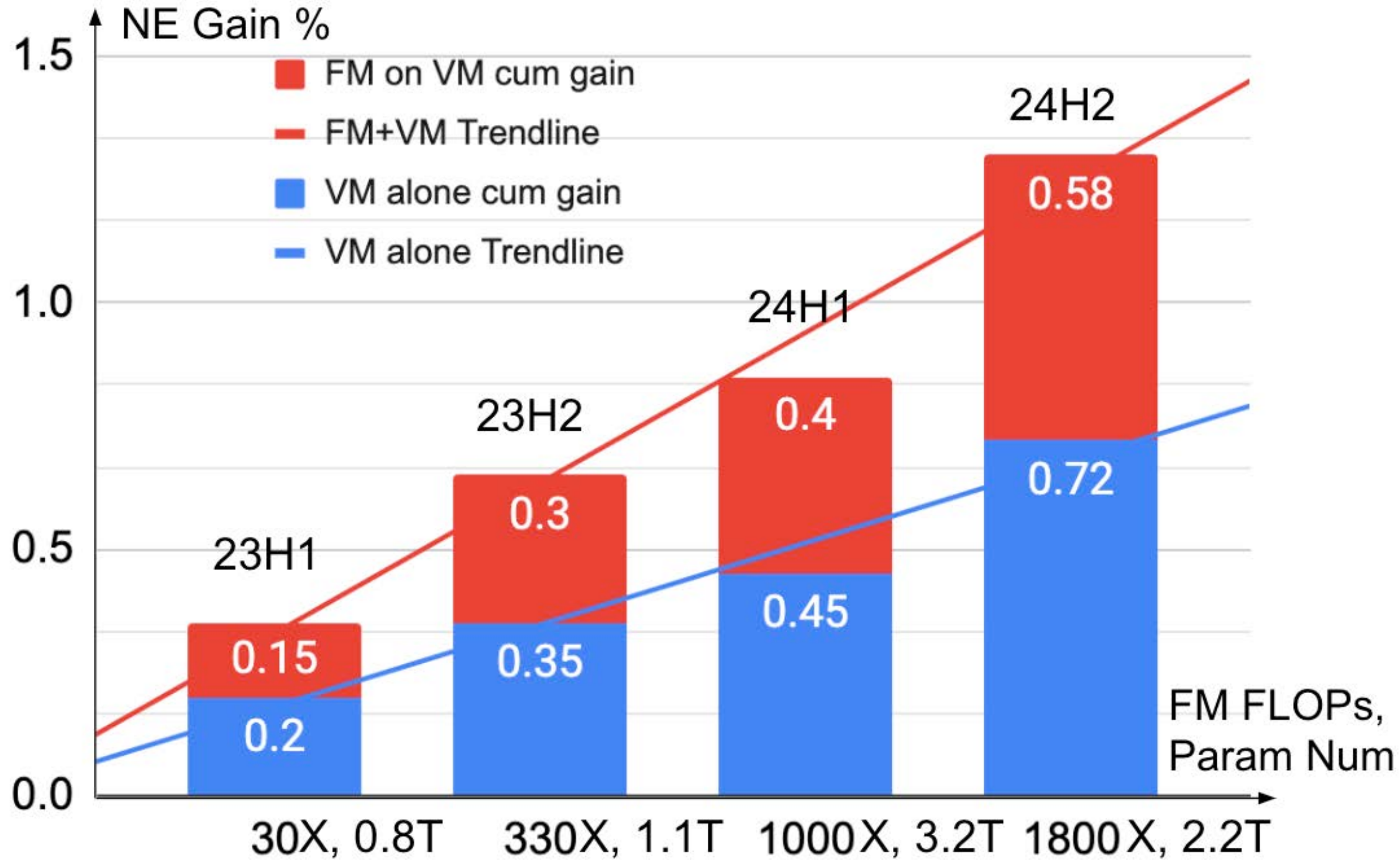}
    \caption{[Internal] Inference NE gain of different-size FM on VM iterations ($1X=60M$ training FLOPs, 1T is 1 Trillion)}
    \label{fig:FM_delta}
\end{figure}

\subsubsection{\textbf{One FM for One VM}}
On internal datasets, we evaluated the impact of different-sized FMs on elevating VM performance. As shown by Figure~\ref{fig:FM_delta}, FM model size ranges from 30X to 1800X where $1X=60M$ training FLOPs. FM model parameter number ranges from 0.8T to 3.2T where 1T is 1 Trillion.
FM scale-up is employing advanced model arches with more features and larger embedding dimensions.
VMs are all 30X and from past halves' real iterations, e.g., 23H1 implies the iteration of the first half in 2023. VMs are iterated by refreshing features and adopting micro model arch changes. 
The baseline of both FMs and VMs is the VM at 22H2.
NE Gain \% refers to the percentage of NE improvement attained vs. baseline. ``VM alone'' NE Gain \% means the NE improvement of VM obtained without FM. Cumulative gain means the number includes the gain since 23H1.
Both FM and VMs are trained with $>300B$ streaming data and inference on the next-day's new arriving data.
Although both VMs and FM are changing from half to half, the value of FM on VM gain is obtained by apple-to-apple comparison.  
We observe that (1) FMs can generate consistent NE gains on VMs from half to half, (2) compared to the trend of the world without FM (blue trendline), the one with FM has steeper slope (red trendline), implying the role of FM on bending the trending curve. 

On public datasets, we evaluated the impact of FM on VMs under different FM-VM pairs in Table~\ref{tab-1: different FM and VM} and \ref{tab-2: different FM and VM}. Data is described in Sec.~\ref{Datasets} and VM is trained on the 5th day with FM's supervision and inference on the 6th day to get the AUC/LogsLoss reading.
Note that the original performance of VMs without FM is limited, because the VMs' model is chosen not to have any SOTA and complex model architectures for the consistency with the problem settings. In the table, ``distill w/ $\mathcal{L}_{\text{ah}}+\mathcal{L}_{\text{sa}}$'' means w/ FM, and ``w/o distill'' means w/o FM.
From the table, we observe consistent AUC and Logloss improvement from having FM across different FM-VM choices, demonstrating that the effectiveness of ExFM can hold in general.

\subsubsection{\textbf{One FM for $N$ VMs}}

\begin{figure}
    \centering
    \includegraphics[width=0.6\linewidth]{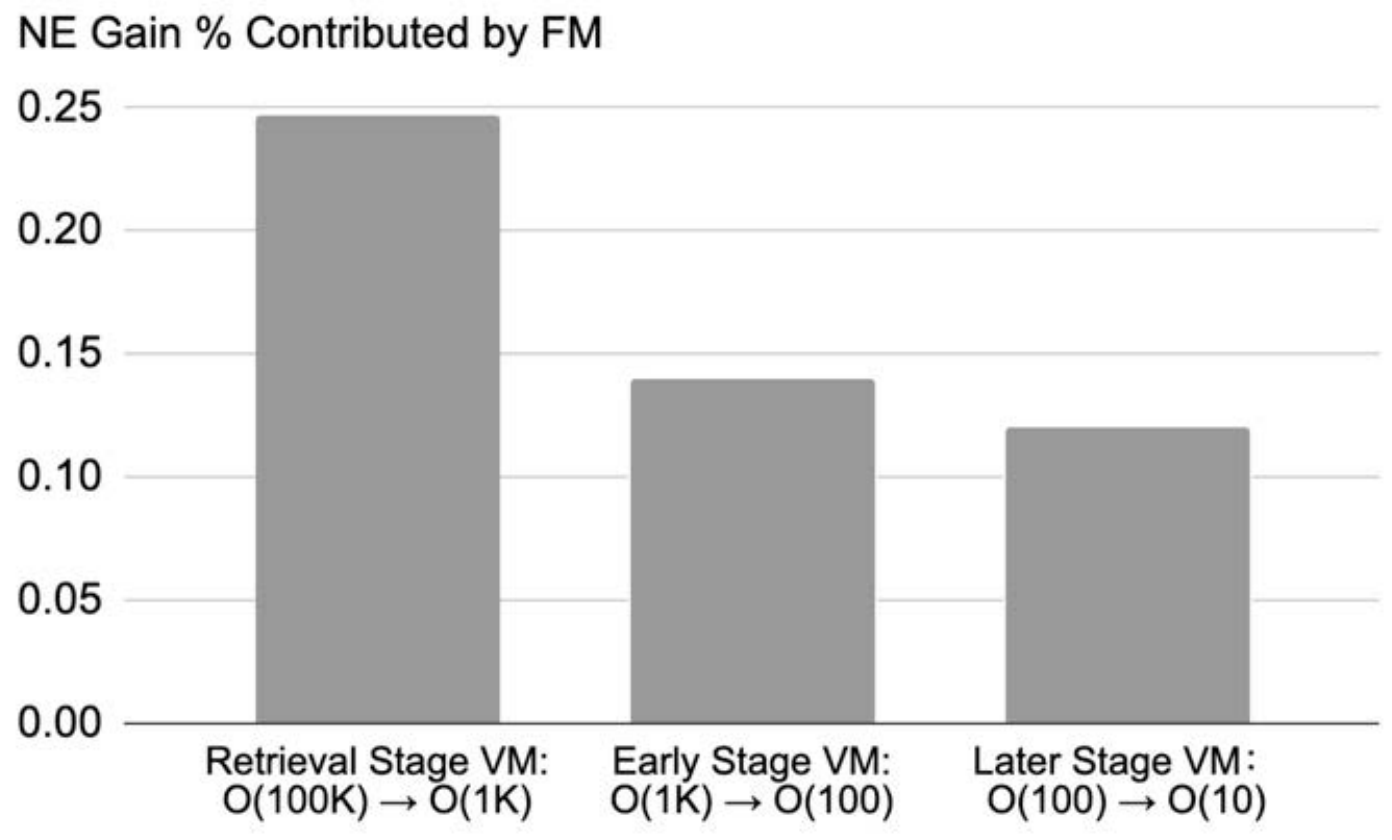}
    \caption{[Internal] Inference NE gain of 1000X, 3.2T FM on cross-stage VMs.} 
    \label{fig: cross-stage VM on internal data}
\end{figure}

\begin{table}[ht]
  \centering
  \begin{minipage}{0.45\textwidth}
    \centering
    \caption{[Public] Impact of FM to cross-domain VMs}
    \begin{tabular}{cccc}
      \toprule
      Methods & Domain & AUC   & logloss \\
      \midrule
      w/o FM & 1     & 0.5025 & 2.2831 \\
      w/ FM  & 1     & \textbf{0.5416} & \textbf{1.8964} \\
      \midrule
      w/o FM & 2     & 0.5024 & 2.5663 \\
      w/ FM & 2     & \textbf{0.5306} & \textbf{1.8812} \\
      \midrule
      w/o FM & 3     & 0.4923 & 3.6295 \\
      w/ FM  & 3     & \textbf{0.5456} & \textbf{1.9003} \\
      \bottomrule
    \end{tabular}%
    \label{tab: multi-domain}%
  \end{minipage}\hfill
  \begin{minipage}{0.45\textwidth}
    \centering
    \caption{[Public] Impact of FM to cross-task VMs}
    \begin{tabular}{cccc}
      \toprule
      Methods & Task  & AUC   & Logloss \\
      \midrule
      w/o FM & Click & 0.7034 & 0.4693 \\
      w/ FM  & Click & \textbf{0.7101} & \textbf{0.4678} \\
      \midrule
      w/o FM & Follow & 0.7613 & 0.00891 \\
      w/ FM  & Follow & \textbf{0.7743} & \textbf{0.00885} \\
      \midrule
      w/o FM & Like  & 0.8531 & 0.01871 \\
      w/ FM  & Like  & \textbf{0.8647} & \textbf{0.01869} \\
      \bottomrule
    \end{tabular}%
    \label{tab: multi-task}%
  \end{minipage}
\end{table}

We evaluate the impact of one FM on multiple VMs across multi-stages in the ads system, including Retrieval, Early, and Later stages, each of which reduces ads candidates, from O(100k) to O(1k), from O(1k) to O(100), and  from O(100) to O(10), respectively. Specicially, we use the 1000X, 3.2T FM in Figure~\ref{fig:FM_delta} and VMs from three stages for experiments on internal dataset.  Retrieval VM and Early stage VM employ the Two-Tower Sparse Network~(TTSN)~\cite{wang2019improving}, and Later stage VM uses DLRM~\cite{naumov2019deep} as the corresponding VM in Figure~\ref{fig:FM_delta}. Training and inference data preparation also follow those of Figure~\ref{fig:FM_delta}. In Figure~\ref{fig: cross-stage VM on internal data}, we observe that the FM brings 0.11\% to 0.25\% additional NE gain to VMs of different stages. The earlier a stage is, the more gains the FM generates. 

To verify the benefits of FM to multiple VMs from different domains, we split the TaobaoAd data into different domains as described in Sec.~\ref{Datasets}. To verify the same thing for different tasks, we use the KuaiVideo dataset where multiple types of feedback are available for prediction. We use the hard-parameter shared DMIN~\cite{navon2024parameter} as the FM and FaM~\cite{rendle2010factorization} dedicatedly trained from scratch for each domain or task as the VM.
Table~\ref{tab: multi-domain} and Table~\ref{tab: multi-task} illustrate the results for different domains and tasks separately. In both tables, we observe that w/ FM outperforms the w/o FM in all tasks and domains, demonstrating the effectiveness of FM for multiple VMs.

\begin{figure}[ht]
    \centering
    \begin{minipage}[b]{0.45\textwidth}
        \centering
        \includegraphics[width=\textwidth]{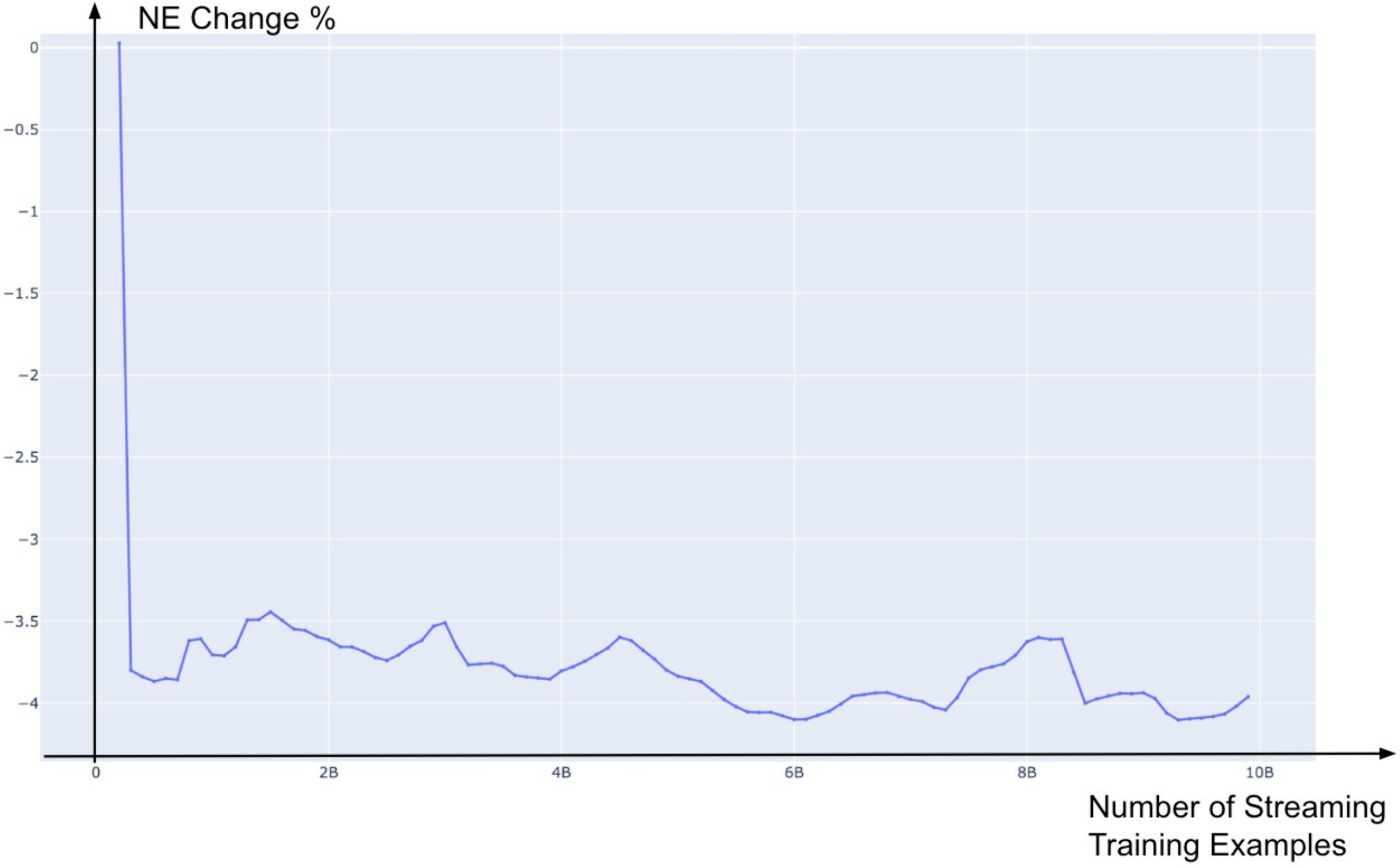}
        \caption{[Internal] NE change of AH with 1000X 3.2T FM}
        \label{fig:ah}
    \end{minipage}
    \hfill
    \begin{minipage}[b]{0.45\textwidth}
        \centering
        \includegraphics[width=\textwidth]{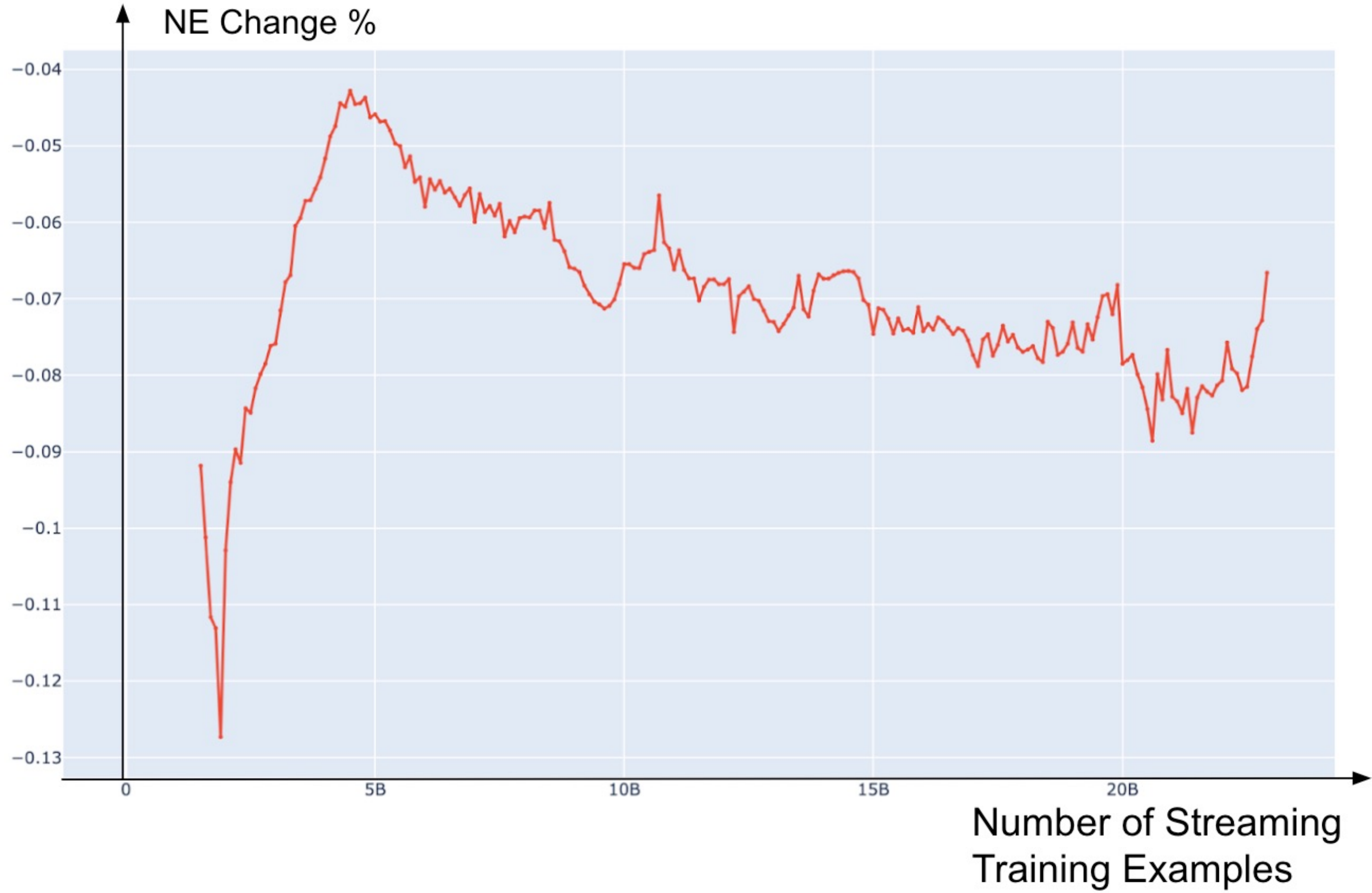}
        \caption{[Internal] NE change of SA with 1800X, 2.2T FM}
        \label{fig:sa}
    \end{minipage}
\end{figure}

\subsection{Effectiveness of of AH and SA}\label{sec:delta of ah/sa}
To understand the role of AH in the success of ExFM, on internal datasets, we compare the VM's NE after including AH~(Eqn.~\ref{auxiliary head loss}) vs. w/o AH (Eqn.~\ref{distillation loss}) under the same 1000X, 3.2T FM and corresponding VM from Figure~\ref{fig:FM_delta}.
Figure~\ref{fig:ah} provides the comparison, where the x-axis is the number of streaming training examples and the value of NE Change at each point is obtained by using the current snapshot to inference on next-batch data, thus equivalent to the inference NE, but in a streaming mode. 
Since NE is the lower the better, the negative `NE Change \%' (-4\%) implies that AH brings a large and stable performance gain compared to baseline. 
Similarly, to understand the role of SA, we compare the VM's NE after including SA vs. w/o SA under the same 1800X, 2.2T FM, and corresponding VM from Figure~\ref{fig:FM_delta}. 
Figure~\ref{fig:sa} shows the comparison, and we observe that with SA, the NE performance of VM on streaming data improves by 0.08\%, with an enlarging trend, which is considered significant.

To reproduce the results on public datasets, we split the dataset as discussed in Sec.~\ref{Datasets} by timestamps to simulate the streaming setting. We apply AH and SA over different pairs of public models on public datasets as described in Section~\ref{exp: baselines}, and the performance of VMs is reported in Table~\ref{tab-1: different FM and VM} and \ref{tab-2: different FM and VM}. Specifically 
\begin{itemize}
    \item distill with $\mathcal{L}_{\text{kd}}$~(Eqn.~\ref{distillation loss}), i.e., the distillation loss
    \item distill with $\mathcal{L}_{\text{ah}}$~(Eqn.~\ref{auxiliary head loss}), i.e., the AH
    \item distill with $\mathcal{L}_{\text{ah}}+\mathcal{L}_{\text{sa}}$~(Eqn.~\ref{student adapter loss for VM}), i.e., AH and SA.
\end{itemize}

\noindent We observe that the AH and SA are consistently superior to the baseline with $\mathcal{L}_{\text{kd}}$ among different pairs of FMs and VMs on various datasets, demonstrating their strong effectiveness in enhancing FM benefits on VMs. 
As the breakdown, it is found that (1) the vanilla distillation w/o AH or SA has very small benefits, e.g., 0.35\% AUC gain for DMIN as FM and FaM as VM on TaobaoAd, (2) the major leap is by AH, e.g., boosting the AUC gain from 0.35\% to 1.11\%, and (3) SA further brings significant gains on top of AH, e.g., lifting the AUC gan from 1.11\% to 1.35\%.

\label{impact of stale FM}
\begin{figure}[htbp]
    \centering
    \includegraphics[width=0.6\textwidth]{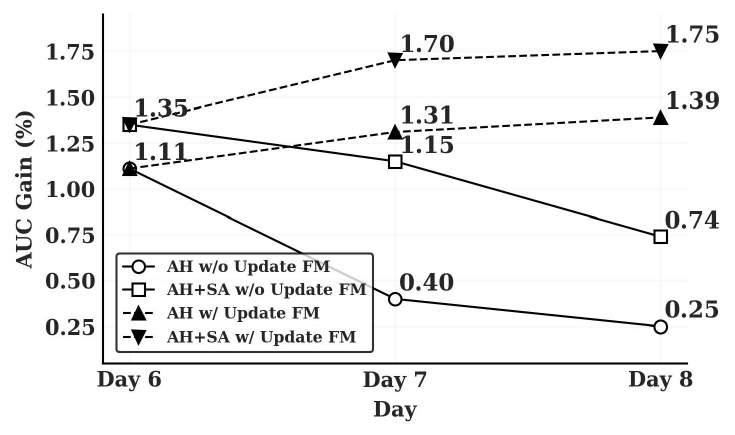}
    \caption{[Public] Impact of SA under FM staleness. }
    \label{fig:staleness-effect}
\end{figure}
\noindent The data points in Table~\ref{tab-1: different FM and VM} and ~\ref{tab-2: different FM and VM} are obtained by one-time inference on the 6th day's data. We are curious about the performance of SA with a bigger freshness gap between FM and VM. To study this, we use DMIN as FM and FaM as VM, stop the FM from being trained on the new day's data, and compare their performance w/ and w/o SA in Figure~\ref{fig:staleness-effect}. We observe that (1) the benefits of FM on VM diminish as the time that FM stops training on new data is longer, (2) SA can generate additional gain no matter whether FM updates or not, but (3) SA cannot revert the diminishing trend when the FM is not updated with new day's data.

\subsection{Impact of hyper-parameters}
\label{sec: parameter analysis}

\begin{figure}[htbp]
    \centering
    \begin{minipage}[b]{0.45\linewidth}
        \centering
        \includegraphics[width=\linewidth]{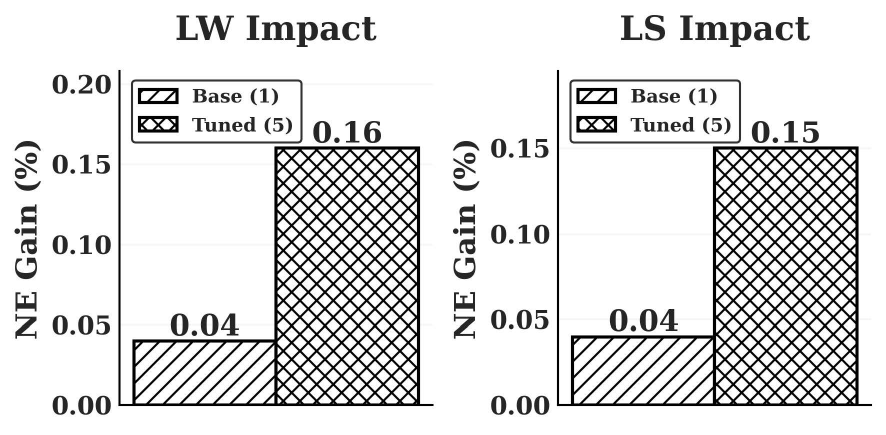}
        \caption{[Internal] Impact of LW and LS on VM's NE}
        \label{fig: tuining}
    \end{minipage}
    \hfill
    \begin{minipage}[b]{0.45\linewidth}
        \centering
        \includegraphics[width=\linewidth]{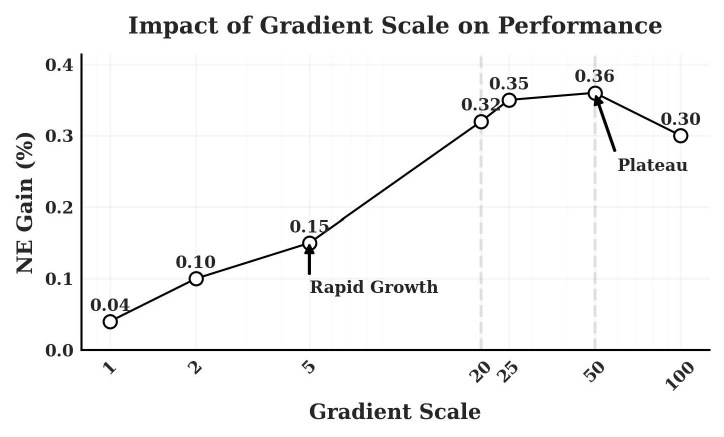}
        \caption{[Internal] Impact of GS on VM's NE}
        \label{fig:grad_scale}
    \end{minipage}
\end{figure}

We notice that the benefits of FM on VM may be sensitive to the choice of GS, LS, and LW as defined in Eqn.~\ref{auxiliary head loss}. On internal datasets, Figure~\ref{fig: tuining} demonstrates the positive impact of increasing LW and LS. We observed that the benefits of FM on VM are improved correspondingly when we increase the value of LW and LS with others fixed. 
Figure~\ref{fig:grad_scale} shows the impact of GS, where there seems to be a sweet spot. The NE gain first increases significantly when GS increases and then tends to plateau when GS is large enough.

\begin{figure}[!htb]
    \centering
    \includegraphics[width=0.7\textwidth]{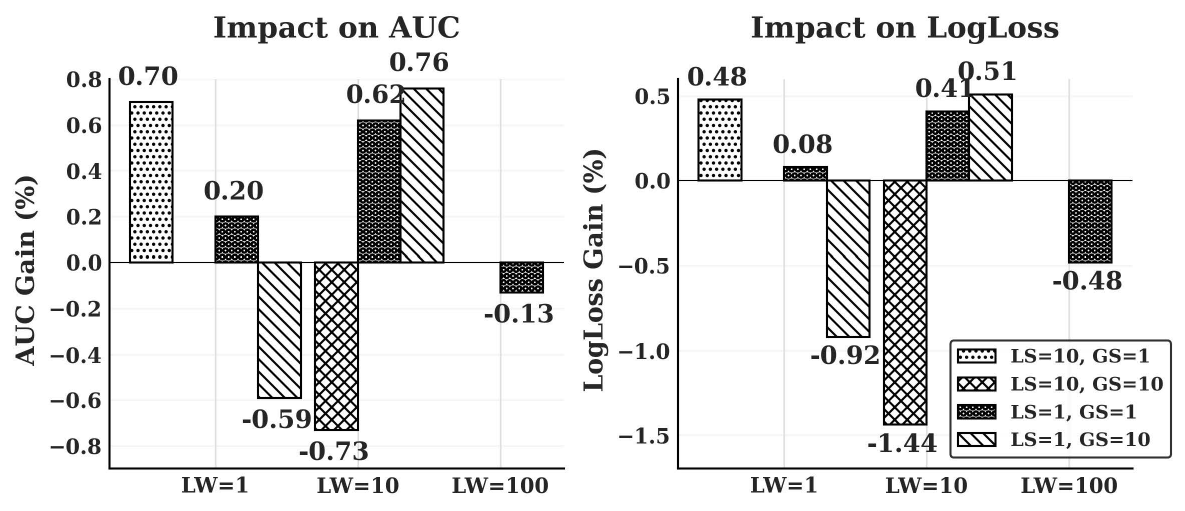}
    \caption{[Public] Joint impact of different factors, i.e., LW, LS, and GS on VM performance.}
    \label{fig:ablation}
\end{figure}

On the internal dataset, the test for the combination effect of LW, LS, and GS is often restricted by the limited training resources. We explore that on the public dataset.
Figure~\ref{fig:ablation} illustrates our experiments on TaobaoAd: (1) The simple combination, i.e., $(LS=1, GS=1, LW=1)$, can only achieve moderate improvement on the VM. (2) Aggressive scaling, e.g., $(LS=10, GS=10, LW=10)$ or $(LS=1, GS=1, LW=100)$, will even be detrimental to the VM's performance. Both observations imply that an appropriate combination of these three scaling techniques could yield significant accuracy improvement and it is important to identify a proper combination for practical usage.

\section{Related Works}
Inspired by the success of large-scale model in NLP~\cite{achiam2023gpt,touvron2023llama}, more and more recent studies have focused on scaling-up recommendation model capacity for better performance~\cite{zhang2024scaling,zhang2024wukong,fang2024scaling,pan2024ads,shin2023scaling,anil2022factory}. 
One fundamental challenge to employ them in industrial-scale applications is the restricted training and inference budget for serving models. 
To overcome that, multiple prior studies explored the adaption of knowledge distillation (KD)~\cite{hinton2015distilling} in recommendation problems, where a large teacher model continuously supervises a compact student model by co-training and finally only the student model is used for serving~\cite{kang2024unbiased,kang2023distillation,chen2023unbiased,liu2022position,kang2021item,kweon2021bidirectional,zhu2020ensembled,tang2018ranking}. 
One limitation of those studies is that they primarily focused on static environments, while in industrial applications, large-volume data continuously arrive. \cite{lee2024continual} proposed to handle that by a continual learning framework. 
Another limitation is that those studies follow co-training based distillation. 
It will increase the training cost of the serving model, as well as the risk of model staleness due to iteration delay, implying performance loss when new incoming data has distribution shifting. \cite{khani2024bridging} proposed to handle that by external distillation where teacher training happens separately from supervising students.

Among existing studies, ours are closest to \cite{lee2024continual} and \cite{khani2024bridging}.
Compared to~\cite{lee2024continual}, our study has the following salient differences: (1) Different settings for the teacher model. To amortize building and maintaining resources, the teacher model in ExFM is an FM that uses an aggregation of student models' feature sets. As a result, the teacher itself often has feature gap and dimension mismatch when compared to an individual VM, so ExFM does not generate student models from the teacher model like~\cite{lee2024continual}. (2) Teacher update does not depend on students. \cite{lee2024continual} instead couples the teacher model’s update with students' by replay learning. Such dependency will increase overhead of teacher update and bring in staleness, as illustrated by Figure~\ref{fig:staleness}, implying a risk of huge performance loss. (3) Consideration of data distribution shifting in an industrial streaming setting. We develop Student Adapter to handle that and provide theoretical guarantees. 
Compared to~\cite{khani2024bridging}, our differences and contributions are as follows. 
Its auxiliary distillation looks similar to our AH, but our AH is an isolated task arch that consumes supervision from FM in a dedicated fashion, not simultaneously consuming true labels like in~\cite{khani2024bridging}. 
\cite{khani2024bridging} lacks theoretical proof to justify its proposal and does not provide a solution to mitigate the distribution gap between FM and VMs due to data distribution shifting. 
Instead, we develop and prove Student Adapter can achieve the goal. 
Moreover, \cite{khani2024bridging} does not contain sufficient details and lacks comprehensive benchmark experiments on external datasets.

\section{Conclusion}
In this paper, we propose the ExFM framework to address two fundamental but overlooked challenges by prior studies on scaling-up Ads recommendation models -- (C1) restricted training and inference budget and (C2) streaming data with distribution shifting.
To overcome C1, ExFM employs the external distillation and the data augmentation service, where teacher training separates from student training and one teacher can supervise multiple VMs, like an FM.
To alleviate the distribution gap between FM and VMs caused by C2, ExFM proposes Auxiliary Head (AH) and Student Adapter (SA), with mathematical guarantees on benefits provided.
ExFM, including its core techniques such as AH and SA, achieved outstanding performance on industrial-scale datasets over multiple tasks and stages. It enabled the serving of a trillion-parameter model without increasing serving latency or negatively impacting user experience, and established the capability to serve LLM-scale ads model in the future. We also experimented on public datasets and were able to reproduce ExFM's outstanding performance, further demonstrating the effectiveness of the proposed framework.

\bibliographystyle{assets/plainnat}
\bibliography{paper}

\clearpage
\newpage
\beginappendix
\section{Appendix}
\subsection{Details of models and training configurations}
\label{Details of datasets, models, and training configurations}
We use the BARS benchmark~\cite{DBLP:conf/sigir/ZhuDSMLCXZ22} and the FuxiCTR~\cite{DBLP:conf/cikm/ZhuLYZH21} to implement all the public models. We follow the default training and model configurations~(e.g., learning rate, training epoch, etc.) in BARS for each public dataset and model used for experiments. 

\subsection{Proof of Theorem~\ref{theo: KD by AH}}
\label{proof of theorem for AH}
We consider a linear model to prove Theorem~\ref{theo: KD by AH} by construction. Let $x \in \mathbb{R}^D$, where $D$ denotes the input dimension. $x$ will be first linearly mapped to a vector of $d \ll D$ dimension by a matrix $Z \in \mathbb{R}^{d \times D}$. We assume that the ground-truth label $y$ is given by $y=u_1^{\top} Z x$, where $u_1 \in \mathbb{R}^d$. For convenience of discussion, let's assume that we have $d-1$ teacher models: for an input $x$, they output $d-1$ soft labels as $w_k^{\top} Z x, k=2, \ldots, d$. 
For simplicity, we assume that $w_k=u_1+\mu u_k, k \in[d]$, with $\mu>0$ being a small weight and $u_1, \ldots, u_d$ form an orthonormal basis for the space of $\mathbb{R}^d$.

\begin{proof}
Let $\left(\alpha_1, \ldots, \alpha_d\right) \in \Delta_{d}$ be the weights to linearly combine the ground-truth labels with teacher labels, where $\Delta$ is a simplex in $\mathbb{R}^d$. 
For a given input $x$, its smoothed label is $\widetilde{y}(x)=\left(u_1+\mu \sum_{k=2}^d \alpha_k u_k\right)^{\top} Z x$.
We fit $\widetilde{y}(x)$ by a two-layer linear model, i.e., $\min _{v \in \mathbb{R}^d, H \in \mathbb{R}^{d \times D}} \frac{1}{T} \sum_{t=1}^{T}\left|v^{\top} H x_t-\widetilde{y}\left(x_t\right)\right|^2$, where $T$ denotes the number of training samples. 
According to the theory of deep linear model~\cite{bartlett2018gradient}, the resulting model, trained by gradient descent, will be able to find a solution with zero loss, i.e., $H^{\top} v=Z^{\top}\left(u_1+\mu \sum_{k=2}^d \alpha_k u_k\right)$, if $\mathrm{E}\left[x x^{\top}\right]$ is non-singular, and we have enough number of training examples. 
The linear prediction model $H^{\top} v$ is different from $Z^{\top} u_1$, and the difference between the two linear prediction models arises from the bias introduced by the label-smoothing.

Now we denote $w_t^1, t=1, \ldots, T$ the solution for the prediction head for the ground-truth labels evolving over time (i.e., each training sample), and by $w_{t}^{k}, k=2, \ldots$ the dynamic solutions for the $d-1$ teacher labels. By the gradient descent method, we have 
\begin{equation}
    \begin{aligned}
Z_{t+1} & =Z_t-2 \eta \sum_{k=1}^d\left(\left\langle w_t^k, Z_t x_t\right\rangle-\left\langle w_k, Z x_t\right\rangle\right) w_t^k x_t^{\top} \\
& =\left(I-2 \eta \sum_{k=1}^d w_t^k\left[w_t^k\right]^{\top}\right) Z_t x_t x_t^{\top}+2 \eta \sum_{k=1}^d w_t^k w_k^{\top} Z x_t x_t^{\top}.
\end{aligned}
\end{equation}
By assuming $x_t \sim \mathcal{N}(0, I)$, $\mathrm{E}_t\left[Z_{t+1}\right]=\left(I-2 \eta \sum_{k=1}^d w_t^k\left[w_t^k\right]^{\top}\right) Z_t+2 \eta \sum_{k=1}^d w_t^k w_k^{\top} Z$. As $w_{t}^{k}$ is updated much more frequently than $Z_{t}$, we have 
\begin{equation}
    w_t^k \approx\left(Z_t Z_t^{\top}\right)^{-1} Z_t Z^{\top} w_k=w_k+\left(Z_t Z_t^{\top}\right)^{-1} Z_t\left(Z-Z_t\right)^{\top} w_k,
\end{equation}
implying that when $Z_{t}$ is close to $Z$, we will have all $w_{t}^{k}$ being close to $w_{t}$. As a result, we have 
\begin{equation}
    \begin{aligned}
      \mathrm{E}_t\left[Z_{t+1}\right] & \approx\left(I-2 \eta \sum_{k=1}^d w_k\left[w_k\right]^{\top}\right) Z_t+2 \eta \sum_{k=1}^d w_k w_k^{\top} Z \\
      & =(1-2 \eta A) Z_t+2 \eta A Z
    \end{aligned}
\end{equation}
where $A=\sum_{k=1}^d w_k w_k^{\top}$ is a full rank matrix, indicating that $Z_{t}$ will converge to $Z$ and thus all $w_{t}^{k}$ will converge to $w_{k}$. This result indicates that by using auxiliary heads, we will guarantee to find the optimal solution $w_{1}$ and $Z$ for predicting the ground-truth labels.
\end{proof}

\subsection{Proof of Theorem~\ref{theo: KD by student adapter}}
\label{proof of theorem 3.2}
We consider the standard regression problem for analysis. 
Let $\left\{x_i \in \mathbb{R}^d, i=1, \ldots, N\right\}$ be the input sampled from a normal distribution $\mathcal{N}(0, I)$. Let the observed output $y_i=x_i^{\top}\left(w+z_i\right)$, where $w$ is the underlying regression model and $z_i \sim \mathcal{N}\left(0, \gamma^2 I\right)$ is an independently sampled vector from a normal distribution that leads to the noise in the observed output value $y_{i}$. In this study, we assume $\gamma \gg 1$. If we learn a regression model directly from the training data, the resulting solution $w_{1}$ is given by 
\begin{equation}
    w_1=\left(\sum_{i=1}^N x_i x_i^{\top}\right)^{-1} \sum_{i=1}^N x_i y_i=w+\left(\sum_{i=1}^N x_i x_i^{\top}\right)^{-1} \sum_{i=1}^N x_i x_i^{\top} z_i.
\end{equation}
Based on the Matrix Chernoff bound, with probability $1-\delta$, we have $\left|\frac{1}{N} \sum_{i=1}^N x_i x_i^{\top}-I\right|_2 \leq O\left(\sqrt{\frac{d}{N} \log \frac{1}{\delta}}\right)$, and $\left|\sum_{i=1}^N x_i x_i^{\top} z_i\right|_2 \leq O\left(\gamma \sqrt{\frac{d}{N} \log \frac{1}{\delta}}\right)$. When $N>\Omega(d \log (1 / \delta))$, we have 
\begin{equation}
    \left|w_1-w\right| \leq O\left(\gamma \sqrt{\frac{d}{N} \log \frac{1}{\delta}}\right),
\end{equation}
implying that to achieve a recovery error $\epsilon$, we need at least $\Omega\left(\frac{\gamma^2 d}{\varepsilon^2} \log \frac{1}{\delta}\right)$ number of training samples. Now we provide the complete proof.
\begin{proof}
As the teacher model will be equipped with the student adapter, i.e., an MLP, to adapt its prediction, we assume the teacher model is parameterized as $\widehat{y}=\widehat{w}^{\top} x+u^{\top} H x$, where $H \in \mathbb{R}^{s \times d}$ is a given orthonormal matrix (i.e., $H_{i, *} H_{j, *}^{\top}=\delta_{i, j}, \forall i, j \in[s]$) with $s \ll d$. For simplicity, we assume that $w-\widehat{w}$ lies in the subspace spanned by the row vectors of $H$, i.e., $w-\widehat{w}=P_H(w-\widehat{w})$, where $P_H=\sum_{i=1}^s H_{i, *}^{\top} H_{i, *}$ is a projection matrix. The overall objective function is given by 
\begin{equation}
    \min _{w^{\prime}, u} \frac{1}{2} \sum_{i=1}^N\left|x_i^{\top} w^{\prime}-y_i\right|^2  +\frac{\alpha}{2} \sum_{i=1}^N\left|x_i^{\top} w^{\prime}-x_i^{\top}\left(\widehat{w} + H^{\top} \operatorname{sg}(u)\right)\right|^2 +\frac{\beta}{2}\left|y_i-x_i^{\top}\left(\widehat{w}+H^{\top} u\right)\right|^2,
\end{equation}
where $\text{sg}$ denotes the stop gradient. We first examine the convergence of $u$, whose expression is given as follows 
\begin{equation}
    \begin{aligned}
u & =\left(\sum_{i=1}^N H x_i x_i^{\top} H^{\top}\right)^{-1} \sum_{i=1}^N\left(w+z_i-\widehat{w}\right)^{\top} x_i H x_i \\
& =H(w-\widehat{w})+\underbrace{\left(\sum_{i=1}^N H x_i x_i^{\top} H^{\top}\right)^{-1} \sum_{i=1}^N H x_i x_i^{\top} z_i}_{:=v}.
\end{aligned}
\end{equation}
Following the standard concentration inequality, with a probability $1-\delta$, we have $|v| \leq O\left(\gamma \sqrt{\frac{s}{N} \log \frac{1}{\delta}}\right)$.
Thus $\left|H^{\top} u-(w-\widehat{w})\right|_2 \leq O\left(\gamma \sqrt{\frac{s}{N} \log \frac{1}{\delta}}\right)$.
Using the result of $u$, we obtain the optimal solution  as $w_{V}=w+\frac{\alpha\left(\widehat{w}+H^{\top} u-w\right)}{1+\alpha}+\frac{1}{1+\alpha}\left(\sum_{i=1}^N x_i x_i^{\top}\right)^{-1} \sum_{i=1}^N x_i x_i^{\top} z_i.$
Following the matrix Bernstein inequality, with an appropriate choice of $\alpha$, we have, with a probability $1-\delta$, 
\begin{equation}
    \left|w_{V}-w\right|_2 \leq O\left(\gamma \sqrt{\frac{(d s)^{1 / 2}}{N} \log \frac{1}{\delta}}\right).
\end{equation}
Compared to the bound for the standard regression model, i.e., $\left|w_1-w\right|_2$, the bound is reduced by the factor $(d / s)^{1 / 4}$.
\end{proof}

\end{document}